\documentclass[ejs]{imsart}

\RequirePackage[OT1]{fontenc}
\RequirePackage{amsthm,amsmath,amssymb,amsfonts,graphicx,subfigure,bm,fullpage}
\RequirePackage[colorlinks,citecolor=blue,urlcolor=blue]{hyperref}
\RequirePackage{hypernat}

\numberwithin{equation}{section}
\theoremstyle{plain}
                          \newtheorem{thm}{Theorem}
                          \newtheorem{lem}[thm]{Lemma}

\theoremstyle{remark}         
\theoremstyle{definition} \newtheorem{defn}{Definition}
                          \newtheorem{exam}{Example}

\newcommand\om{\Omega}
\newcommand\real{\mathbb{R}}
\newcommand\A{\mathcal{A}}

\newcommand\bx{\bm{x}}
\newcommand\by{\bm{y}}

\newcommand\bX{\bm{X}}
\newcommand\bY{\bm{Y}}

\newcommand\balp{\bm\alpha}
\newcommand\blam{\bm\lambda}

\newcommand\R{\mathcal{R}}

\def\woMR#1{\w@MR#1MR#1MR\relax}%
\def\w@MR#1MR#2MR#3\relax{#2}

\def\@MR#1 #2\relax#3{%
 \href{http://www.ams.org/mathscinet-getitem?mr=#1}%
 {\MRfixed{#3}}}%

\def\MRfixed{MR\woMR}%

\usepackage{varioref}		    
\labelformat{section}{Section~#1}   
\labelformat{figure}{Figure~#1}
\labelformat{table}{Table~#1}

\begin{document}
\begin{frontmatter}
\title{Recursive partitioning and multi-scale modeling on conditional densities}
\runtitle{A multi-scale prior for conditional distributions}
\author{\fnms{Li} \snm{Ma} \thanksref{t1}\ead[label=e1]{li.ma@duke.edu}}
\affiliation{Duke University}
\address{Department of Statistical Science \\ Duke University  \\ Durham, NC 27708-0251, USA \\ \printead{e1}}
\runauthor{L. Ma}
\date{August 27th, 2011}

\thankstext{t1}{Supported in part by NSF grants DMS-1309057 and DMS-1612889, and a Google Faculty Research Award.}


\begin{abstract}
We introduce a nonparametric prior on the conditional distribution of a (univariate or multivariate) response given a set of predictors. The prior is constructed in the form of a two-stage generative procedure, which in the first stage recursively partitions the predictor space, and then in the second stage generates the conditional distribution by a multi-scale nonparametric density model on each predictor partition block generated in the first stage. This design allows adaptive smoothing on both the predictor space and the response space, and it results in the full posterior conjugacy of the model, allowing exact Bayesian inference to be completed analytically through a forward-backward recursive algorithm without the need of MCMC, and thus enjoying high computational efficiency (scaling linearly with the sample size). We show that this prior enjoys desirable theoretical properties such as full $L_1$ support and posterior consistency. We illustrate how to apply the model to a variety of inference problems such as conditional density estimation as well as hypothesis testing and model selection in a manner similar to applying a parametric conjugate prior, while attaining full nonparametricity. Also provided is a comparison to two other state-of-the-art Bayesian nonparametric models for conditional densities in both model fit and computational time. A real data example from flow cytometry containing 455,472 observations is given to illustrate the substantial computational efficiency of our method and its application to multivariate problems.
\end{abstract}
 


\end{frontmatter}

\section{Introduction}
\label{sec:introduction}
In recent years there has been growing interest in nonparametrically modeling probability densities based on multi-scale partitioning of the sample space. A prime example in the Bayesian nonparametric literature is the P\'olya tree (PT) \cite{ferguson:1973,lavine:1992,mauldin:1992} and its extensions \cite{hanson:2002,hanson:2006,wongandma:2010,jara:2011,ma:2016}. In particular, Wong and Ma \cite{wongandma:2010} introduced randomization into the partitioning component (involving both random selection of partition directions as well as optional stopping) of the PT framework, which enhances the model's ability to approximate the shape and smoothness of the underlying density. A PT model with these features is called an optional P\'olya tree (OPT).

A further desirable feature of the PT and its relatives such as the OPT and the more recently introduced adaptive P\'olya tree (APT) \cite{ma:2016} is the computational ease for carrying out inference. In turns out that the extra component of randomized partitioning such as that employed in the OPT does not impair the conjugacy enjoyed by the PT. For example, after observing i.i.d.\ data, the corresponding posterior of an OPT is still an OPT, that is, the same generative procedure for random probability distributions with its parameters updated to their posterior
values. Moreover, the corresponding posterior parameter values can be
computed {\em exactly} through a sequence of recursive computations, which is in essence a forward-backward algorithm \cite{liu:2001}. This, together
with the constructive nature of these models, allows one
to draw samples from the exact posterior directly without resorting to
Markov Chain Monte Carlo (MCMC) procedures, and to compute various summary statistics of the posterior analytically. Furthermore, the marginal posterior of the random partitioning adapts to
the underlying structure of the data---the sample space will with high posterior probability be more finely divided in places where the underlying distribution has richer structure, i.e. less uniform topological shape.

Motivated by the computational efficiency and statistical properties of the OPT, which is tied to its use of recursive random partitioning, we aim to further exploit the random recursive partitioning idea in the context of multi-scale density modeling, and build such a model for {\em conditional} densities for a response (vector) $\bY$ given a predictor (vector) $\bX$. The objective is to construct a flexible nonparametric model for conditional distributions that maintain all of the desirable statistical and computational properties of PT and OPT.

A variety of inference tasks involve the estimation, prediction, and testing regarding conditional distributions, and nonparametric inference on conditional densities has been studied from both
frequentist and Bayesian perspectives. Many frequentist works
are based on kernel estimation methods \cite{fan:1996,hall:1999,fan:2004}, and they achieve proper smoothing through bandwidth selection, which often involves resampling procedures such as cross-validation
\cite{bashtannyk:2001,hyndman:2001,fan:2004} and the bootstrap
\cite{hall:1999}. An alternative frequentist strategy introduced more recently is to employ the so-called block-wise shrinkage \cite{efromovich:2007,efromovich:2010}. In Bayesian nonparametrics, inference on conditional distributions is often referred to as covariate-dependent distribution modeling, and existing methods fall into two categories. The first category is methods that construct priors for the joint distribution of the
response and the predictors, and then use the induced conditional
distribution for inference. Some examples are
\cite{muller:1996,roeder:1997,norets:2012,taddy:2010}, which propose using
mixtures of multivariate normals as the model for joint
distributions, along with different priors for the mixing distribution. The
other category is methods that construct conditional
distributions directly without specifying the marginal distribution of the predictors.  Many of these methods are
based on extending the stick breaking construction for the Dirichlet Process
(DP) \cite{sethuraman:1994}. Some notable examples, among others, are proposed in
\cite{maceachern:1999,deiorio:2004,gelfand:2005,griffin:2006,dunson:2008,
 chung:2009,rodriguez:2011, barrientos:2017}. Some recent works in this category do not utilize stick breaking. In
\cite{tokdar:2010}, the authors propose to use the logistic
Gaussian process \cite{lenk:1998,tokdar:2007} together with subspace projection to construct smooth
conditional distributions. In \cite{jara:2011}, the authors incorporate covariate dependency into tail-free processes by generating the
conditional tail probabilities from covariate-dependent logistic Gaussian
processes, and propose a mixture of such processes as
a way for modeling conditional distributions. The authors of \cite{lijoi:2014} introduce dependent normalized complete random measures. In \cite{trippa:2011} the authors introduce the covariate-dependent multivariate Beta process, and use it to generate the conditional tail probabilities of P\'olya trees. More recently, in \cite{shen:2016} the authors use the tensor product of B-splines to construct a prior for conditional densities, and incorporate a variable selection feature. 
While many of these nonparametric models on conditional distributions enjoy desirable theoretical properties, inference using these priors 
generally relies on intense MCMC sampling, and can take substantial computing time even when both the response and the covariate are one-dimensional.

We introduce a new prior, called the conditional optional P\'olya tree, for the conditional density of $\bY$ given $\bX$, in the form of a two-stage generative procedure consisting of first randomly partitioning the {\em predictor} space $\om_{\bX}$, and then for each predictor partition block, generates the response distribution on each block using an OPT, which implicitly employs a further random partitioning of the {\em response} space $\om_{\bY}$. We show that this new prior is a fully nonparametric model and yet achieves extremely high computational efficiency even for multivariate responses and covariates. It enjoys all of the desirable theoretical properties of the PT and the OPT
priors---namely large support, posterior consistency, and posterior conjugacy, and its posterior parameters can also be computed exactly through forward-backward recursion. Under this two-stage design, the posterior distribution on the partitions reflect the
structure of the conditional distribution at two levels---first, the predictor space will be partitioned
finely in parts where the conditional distribution changes most
abruptly, shedding light on how the conditional distribution depends
on the predictors; second, the response space will be divided adaptively for different locations of the predictor space, to capture the local structure of the conditional density through adaptive smoothing.

The rest of the paper is organized as follows. In
\ref{sec:cond_opt} we introduce our two-stage prior and show that it is fully nonparametric---with full (integrated) $L_1$ support---for conditional densities. In addition, we make a connection to Bayesian CART and show that our method can be considered a nonparametric version of the latter. In \ref{sec:bayes_inference} we show the full conjugacy of the model, derive the exact form of the posterior through forward-backward recursion, and thereby provide a recipe for carrying out Bayesian inference using the prior. We also prove the posterior consistency of such inference.
In \ref{sec:comp_issues} we discuss practical computational issues in implementing the inference. In
\ref{sec:examples} we provide four simulation examples to illustrate 
the work of our method. The first two are for estimating conditional
densities, and the last two concern model selection and hypothesis testing. In \ref{sec:flow} we apply the proposed method to estimating conditional densities in a flow cytometry data set involving a large number (455,472) of observations, and demonstrate the computational efficiency of the method and its application when both the response and the predictor are multivariate. \ref{sec:discussion} concludes with some discussions. All proofs are given in the Appendix.

\section{Conditional optional P\'olya trees}
\label{sec:cond_opt}
In this section we introduce our proposed prior constructively in terms of a two-stage generative procedure that produces random conditional densities. First we introduce some notions and notations that will be used throughout. Let each observation be a predictor-response pair $(\bX,\bY)$, where $\bX$ denotes the predictor (or covariate) vector and $\bY$ the response (vector) with $\om_{\bX}$ being the
predictor space and $\om_{\bY}$ the response space. In this work we consider sample spaces that are either finite spaces, compact Euclidean rectangles, or a product of the two, and $\om_{\bX}$ and $\om_{\bY}$ do not
have to be of the same type. (See for instance Example~\ref{ex:bernoulli}.) Let
$\mu_{\bX}$ and $\mu_{\bY}$ be the ``natural'' measures on 
$\om_{\bX}$ and $\om_{\bY}$. (That is, the counting measure for finite spaces,
the Lebesgue measure for Euclidean rectangles, and the corresponding product measure if the space is a product of the two.) Let $\mu = \mu_{\bX}
\times \mu_{\bY}$ be the ``natural'' product measure on the joint sample space $\om_{\bX}\times \om_{\bY}$. 

A {\em partition rule} $\R$ on a sample space $\om$ specifies a collection of possible ways to divide any subset $A$ of $\om$ into a number of smaller sets.
For example, for $\om=[0,1]^k$, the unit rectangle in $\real^k$, the {\em coordinate-wise dyadic mid-split rule} allows each rectangular subset $A$ of $\om$ whose sides are parallel to the $k$ coordinates to be divided into two halves at the middle of the range of each coordinate. 
For simplicity, in this work we only consider partition rules that allow a {\em finite} number of ways for dividing each set. Such partition rules are said to be {\em finite}. 
(Interested readers can refer to \cite[Sec.~2]{maandwong:2011} for a more detailed treatment of partition rules and to Examples~1 and 2 in \cite{wongandma:2010} for examples of the coordinate-wise dyadic mid-split rule over Euclidean rectangles and $2^k$ contingency tables.)

We are now ready to introduce our prior for conditional densities as a two-stage constructive procedure. It is important to note that the following describes the generation of conditional densities under our prior and not the operational steps for inference under the prior, which will be addressed \ref{sec:bayes_inference} and \ref{sec:comp_issues}.
\vspace{0.7em}

\noindent {\bf Stage I. Predictor partition:} We randomly partition $\om_{\bX}$ according to a given partition rule $\R_{\bX}$ on $\om_{\bX}$ in the following recursive manner. 
Starting from $A=\om_{\bX}$, draw a Bernoulli variable 
\[
S(A) \sim {\rm Bernoulli}(\rho(A)).
\]
That is, ${\rm P}(S(A)=1)=\rho(A)$. If $S(A)=1$, then the partitioning procedure on $A$ terminates and we arrive at a trivial partition of a single block over $A$. (Thus $S(A)$ is called the stopping variable, and $\rho(A)$ the stopping probability.) If instead $S(A)=0$, then we randomly select one out of the possible ways for dividing $A$ under $\R_{\bX}$ and partition $A$ accordingly. More specifically, if there are $N(A)$ ways to divide $A$ under $\R_{\bX}$, we randomly draw 
\[ J(A) \in \{1,2,\ldots,N(A)\} \text{ such that ${\rm P}(J(A)=j)=\lambda_j(A)$ for
$j=1,2,\ldots,N(A)$ with $\sum_{j=1}^{N(A)} \lambda_j(A)=1$}\]
 and partition $A$ in the $j$th
way if $J(A)=j$. (We call $\blam(A)=\bigl(\lambda_1(A),\lambda_2(A),\ldots,\lambda_{N(A)}(A)\bigr)$ the partition selection probabilities for $A$.) Let $K^{j}(A)$ be the number of child sets that arise from this partition, and let $A^{j}_{1}, A^{j}_{2}, \ldots,
A^{j}_{K(A)}$ denote these children. We then repeat the same partition procedure, starting from the drawing of a stopping variable, on each of these children. 

The following lemma, first proved in \cite{wongandma:2010}, states that as long as the stopping probabilities are (uniformly) away from 0, this random recursive partitioning procedure will eventually terminate almost everywhere and produce a well-defined partition of $\om_{\bX}$. 
\begin{lem}
\label{lem:exist}
If there exists a $\delta >0$ such that the
stopping probability $\rho(A)>\delta$ for all $A\subset \om_{\bX}$ that could arise after a finite number of levels of recursive partition, then with probability 1
the recursive partition procedure on $\om_{\bX}$ will stop $\mu_{\bX}$ a.e. 
\end{lem}
\vspace{0.5em}

\noindent {\bf Stage II. Generating conditional densities:}
Next we move onto the second stage of the procedure to generate the conditional density of the response $\bY$ on each of the predictor partition blocks generated in Stage~I. Specifically, for each stopped subset $A$ on $\om_{\bX}$ produced in Stage~I, we let the conditional distribution
of $\bY$ given $\bX=x$ be the same across all $x\in A$, and generate this (conditional)
distribution on $\om_{\bY}$, denoted as $q_{\bY}^{0,A}$, from a ``local''
prior.

When the response space $\om_{\bY}$ is finite, $q_{\bY}^{0,A}$ is simply a multinomial distribution, and so a simple choice of such a local prior is the Dirichlet prior: $q_{\bY}^{0,A} \sim {\rm Dirichlet}(\balp^A_{\bY})$
where $\balp^A_{\bY}$ represents the {\em pseudo-count} hyperparameters of the Dirichlet. In this case, we note that the two-stage prior essentially reduces to a version of the Bayesian CART proposed by Chipman et al in \cite{chipman:1998} for the classification problem. 
When $\om_{\bY}$ is infinite (or finite but with a large number of elements), one may restrict $q_{\bY}^{0,A}$ to be from a parametric family. For example, when $\om_{\bY}=\real$, one may require $q^{0,A}_{\bY}$ to be normal with some mean $\mu_{A}$ and variance $\sigma_{A}^2$, and let
$\mu_{A}|\sigma_A^2 \sim {\rm N}(\mu_0,\sigma^2)$ and $\sigma_A^2 \sim \text{inverse-Gamma}(\nu/2,\nu\kappa/2)$.
In this case the two-stage prior again reduces to a Bayesian CART, this time for the regression problem \cite{chipman:1998}. 

The focus of our current work, however, is on the case when no parametric assumptions are placed on the conditional density. To this end, one can draw $q^{0,A}_{\bY}$ from a nonparametric prior. A desirable choice for the local prior, which will result in analytic simplicity and computational efficiency as we will later show, is a P\'olya tree type model \cite{ma:2016}, and in particular an optional P\'olya tree (OPT) distribution~\cite{wongandma:2010}:
\[
q_{\bY}^{0,A} \sim {\rm OPT}(\R_{\bY}^{A};\rho_{\bY}^{A},\blam_{\bY}^{A},\balp_{\bY}^{A})
\] 
independently across $A$s given the partition, where $\R_{\bY}^A$  denotes a partition rule on $\om_{\bY}$ and $\rho_{\bY}^{A}$, $\blam_{\bY}^{A}$,
and $\balp_{\bY}^{A}$ are respectively the stopping, selection, and pseudo-count hyperparameters \cite{wongandma:2010}. In general we allow the partition rule for these ``local'' OPTs to
depend on $A$ as indicated in the superscript, but adopting a common
partition rule on $\om_{\bY}$---that is to let $\R_{\bY}^{A}\equiv
\R_{\bY}$ for all $A$---will suffice for most problems. In the rest of the paper, unless stated otherwise we assume that a common rule $\R_{\bY}$ is adopted. 

This completes the description of our two-stage procedure. We now formally define the resulting prior.
\begin{defn}
A conditional distribution that arises from the above
two-stage procedure is said to have a {\em conditional optional P\'olya tree} (cond-OPT) distribution. The hyperparameters are the predictor partition rule
$\R_{\bX}$, the response partition rule $\R_{\bY}$, the stopping probability $\rho(A)$, the partition selection probabilities
$\blam(A)$, and the local parameters $(\rho_{\bY}^{A},\blam_{\bY}^{A},\balp_{\bY}^{A})$ for all $A\subset\om_{\bX}$ that could arise during the predictor partition under $\R_{\bX}$.
\end{defn}   
\noindent Remark I: To ensure that this definition is meaningful, one must check that the two-stage procedure will in fact generate a well-defined conditional distribution with probability 1. To see this, first note that because the collection of all potential sets $A$ on $\om_{\bX}$
that can arise during Stage~I is countable, by Theorem~1 in \cite{wongandma:2010}, with
probability~1, the two-stage procedure will generate an absolutely continuous conditional distribution
of $\bY$ given $\bX=x$ for $x$ in the stopped part of $\om_{\bX}$, provided that $\rho^A_{\bY}$ is uniformly away from 0. The two-stage generation procedure for the conditional density of $\bY$ can then be completed by
letting $\bY$ given $\bX$ be uniform on $\om_{\bY}$
for the $\mu_{\bX}$-null subset of $\om_{\bX}$ on which the recursive partition in Stage~I
never stops. 
\vspace{0.5em}

\noindent Remark II: While the cond-OPT prior involves many hyperparameters, one can appeal to very simple symmetry and self-similarity principles for choosing their values. Specifically, such considerations lead to the simple choice: (i) $\rho(A)\equiv \rho\in [0,1]$, (ii) $\lambda_j(A)=1/N(A)$, and (iii) $\rho_{\bY}^A\equiv \rho_{\bY}$, $\blam_{\bY}^A\equiv \blam_{\bY}$, and $\balp_{\bY}^A \equiv \balp_{\bY}$ for all $A$, following the default choices in \cite{wongandma:2010}. 
We note that when useful prior knowledge about the structure of the underlying distribution is not available or when one is unwilling to assume particular structure over the distribution, it is desirable to specify the prior parameters in a symmetric and self-similar way. The common stopping probability $\rho$ should not be too close to 0 or 1, but taking a moderate value between 0.1 and 0.9. A sensitivity analysis for such choices demonstrating the robustness of such choices in the context of OPTs is provided in \cite{maandwong:2011}. As for the partition rules, the coordinate-wise dyadic mid-split rule can serve as a simple default choice for both $\R_{\bX}$ and $\R_{\bY}$. We will adopt such a specification in all of our numerical examples. 
\vspace{0.5em}

\noindent Remark III: One constraint in the cond-OPT is that given the random partition generated in Stage~I, the generation of the conditional distribution across different predictor blocks is independent, i.e., in a similar manner as that for Bayesian CART. As we shall see, this constraint is key to the tremendous computational efficiency of the model. It is important to note however that due to the randomized partitioning incurred in Stage~I,
the marginal prior for the conditional distributions on nearby values of $\bX$ are in fact dependent, thereby achieving smoothing over $\om_{\bX}$ to some extent. More flexible smoothing could be achieved through modeling the ``local'' priors jointly, but that would incur the need for MCMC sampling and the most desirable feature of PT type models would be lost.
\vspace{0.5em}

We have emphasized that the cond-OPT prior imposes no parametric assumptions on the conditional distribution. One may wonder whether this prior is truly ``nonparametric'' in the sense that it can generate all possible conditional densities. Our next theorem confirms this---under mild conditions on the parameters, which the default specification satisfies, the cond-OPT will
place positive probability in arbitrarily small $L_1$ neighborhoods
of any conditional density. (A definition of an $L_1$
neighborhood for conditional densities is also implied in the statement of the theorem.)
\begin{thm}[Large support]
\label{thm:large_support}
Suppose $q(\cdot|\cdot)$ is a conditional density function that arises
from a cond-OPT prior whose parameters $\rho(A)$ and $\blam(A)$ for all $A$ that could arise during the recursive partitioning on $\om_{\bX}$ are uniformly away from 0 and 1, and the local OPTs all have full $L_1$ support on the densities on $\om_{\bY}$. Moreover, suppose that the underlying partition rules $\R_{\bX}$ and $\R_{\bY}$ both satisfy the following ``fine partition
criterion'': $\forall \epsilon>0$, there exists a partition of the corresponding sample space such that the diameter of
each partition block is less than $\epsilon$. Then for any conditional density function $f(\cdot|\cdot) :
\om_{\bY} \times \om_{\bX} 
\rightarrow [0,\infty)$, and any $\tau >0$, 
\[
P\left( \int |q(y|x) - f(y|x)| \mu(dx \times dy) < \tau \right) > 0.
\]
Furthermore, let
$f_{\bX}(x)$ be any density function on $\om_{\bX}$ w.r.t. $\mu_{\bX}$. Then we have $\forall \tau>0$,
\[
P\left( \int |q(y|x) - f(y|x)| f_{\bX}(x) \mu(dx \times dy) < \tau \right) > 0.
\]
\end{thm}
\noindent Remark: Sufficient conditions for OPTs to have full $L_1$ support on densities is given in Theorem~2 of $\cite{wongandma:2010}$.

\section{Bayesian inference with cond-OPT}
\label{sec:bayes_inference}
Next we investigate how Bayesian inference on conditional densities
can be carried out using this prior. First, we note that Chipman et al \cite{chipman:1998} and Denison et al \cite{denison:1998} each proposed MCMC algorithms that enable posterior inference for Bayesian CART. These sampling and stochastic search algorithms can be applied directly here as the local OPT priors can be marginalized out and so the marginal likelihood under each partition tree that arises in Stage~I of the cond-OPT is available in closed form \cite{wongandma:2010,maandwong:2011}. However, as noted in \cite{chipman:1998} and other works, due to the multi-modal nature of tree structured models, the mixing behavior of the MCMC algorithms is often undesirable. This problem is exacerbated in higher dimensional settings. Chipman et al \cite{chipman:1998} suggested using MCMC as a tool for searching for good models rather than a reliable way of sampling from the actual posterior. 

The main result of this section is that under simple partition rules such as the coordinate-wise dyadic mid-split rule, Bayesian inference under a cond-OPT prior can be carried out in an {\em exact} manner in the sense that the corresponding posterior distribution can be computed in closed form and directly sampled from, without resorting to MCMC algorithms. Not only is the computation feasible for multivariate sample spaces of moderate dimensions, but it is in fact highly efficient, scaling linearly with the number of observations.

First let us investigate what the posterior of a cond-OPT prior is. Suppose we have observed $(\bx,\by)=\{(x_1,y_1),(x_2,y_2),\ldots,(x_n,y_n)\}$ where given the $x_i$'s, the $y_i$'s are independent with some density $q(y|x)$. We assume that $q(\cdot|\cdot)$ has a cond-OPT prior denoted by $\pi$. 
Further, for any $A \subset \om_{\bX}$ we let 
\[\bx(A):=\{x_1,x_2,\ldots,x_n\}\cap A \quad \text{and} \quad \by(A) := \{y_i: x_i \in A, i=1,2,\ldots,n\},\]
and let $n(A)$ denote the number of observations with predictors lying in $A$, that is $n(A)=|\bx(A)|=|\by(A)|$.

For $A\subset \om_{\bX}$, we use $q(A)$ to denote the (conditional) likelihood under $q(\cdot|\cdot)$ contributed from the data with predictors $x \in A$. That is
\[
q(A) := \prod_{i:x_i \in A} q(y_i|x_i).
\]
Then conditional on the event that $A$ arises during the recursive partition procedure on $\om_{\bX}$, we can write $q(A)$ recursively in terms of $S(A)$, $J(A)$, and $q^A_{\bY}$ as follows
\[
q(A) = \left\{ \begin{array}{ll} q^{0}(A) & \text{ if $S(A)=1$}\\\\
\prod_{i=1}^{K^{j}(A)}q(A^{j}_{i}) & \text{ if $S(A)=0$ and $J(A)=j$,}    \end{array}
\right.
\]
where 
\[ 
q^{0}(A) := \prod_{i:x_i\in A} q_{\bY}^{0,A}(y_i),
\]
the likelihood from the data with $x\in A$ if the partitioning stops on $A$. Equivalently, we can write
\begin{align}
q(A) &=
S(A)q^{0}(A)+(1-S(A))\prod_{i=1}^{K^{J(A)}(A)}
q(A^{J(A)}_i).
\label{eq:cond}
\end{align}
Integrating out the randomness over both sides of Eq.~\eqref{eq:cond}, we
get
\begin{align}
\Phi(A)=\rho(A)M(A) +
\bigl(1-\rho(A)\bigr)\sum_{j=1}^{N(A)}\lambda_j(A) \prod_i \Phi(A^j_i),
\label{eq:cond_int}
\end{align}
where 
\[\Phi(A):=
\int q(A)\pi(dq\,|\,\text{A arises during the recursive partitioning})
\]
is defined to be the marginal likelihood from data with $x \in A$ given that $A$ arises during the recursive partitioning on $\om_{\bX}$, whereas 
\begin{align}
\label{eq:m}
M(A):=\int q^{0}(A)\pi(dq_{\bY}^{0,A})
\end{align}
is the marginal likelihood from the data with $x \in A$ if the recursive partitioning procedure stops on $A$ and the integration is taken over the local OPT$(\R_{\bY};\rho_{\bY}^A,\blam_{\bY}^A,\balp_{\bY}^A)$ prior for $q_{\bY}^{0,A}$. We note that Eqs.~\eqref{eq:cond}, \eqref{eq:cond_int} and \eqref{eq:m} hold for Bayesian CART as well, with $M(A)$ being the corresponding marginal likelihood of the local normal model or the multinomial model under the corresponding priors such as those given earlier.

Eq.~\eqref{eq:cond_int} provides a recursive recipe for calculating $\Phi(A)$ for all $A$. It is recursive in the sense that $\Phi(A)$ is computed based on the value of $\Phi(\cdot)$ on $A$'s children. (Of course, to complete the calculation the recursion must eventually terminate everywhere on $\om_{\bX}$. We shall describe the terminal conditions in the next section.) This recursive algorithm is a special case of the forward-backward algorithm \cite{ma:2016}. 

The next theorem establishes the posterior conjugacy of cond-OPT.
\begin{thm}[Conjugacy]
\label{thm:conjugacy}
After observing $\{(x_1,y_1),(x_2,y_2),\ldots,(x_n,y_n)\}$ where given the $x_i$'s, the $y_i$'s are independent with density $q(y|x)$, which has a cond-OPT prior, the posterior of
$q(\cdot|\cdot)$ is again a cond-OPT (with the same partition rules on $\om_{\bX}$ and $\om_{\bY}$ as the prior). Moreover,
for each $A\subset \om_{\bX}$ that could arise during the recursive partitioning, the posterior parameters are given as follows.
\begin{enumerate}
\item Stopping probability:
  \[ \rho(A|\bx,\by)=\rho(A)M(A)/\Phi(A). \]
\item Selection probabilities: 
\[\lambda_j(A|\bx,\by) = \lambda_j(A)\frac{(1-\rho(A))\prod_{i=1}^{K^j(A)} \Phi(A^j_i)}{\Phi(A)-\rho(A) M(A)}. \]
\item The local parameters:
$\tilde{\rho}_{\bY}^{A}$, $\tilde{\blam}_{\bY}^{A}$, and $\tilde{\balp}_{\bY}^{A}$ are
the corresponding posterior parameters for
the local OPT after updating using the observed values for the response $\by(A)$, ${\rm OPT}(\R^A_{\bY};\tilde{\rho}_{\bY}^{A},\tilde{\blam}_{\bY}^{A},\tilde{\balp}_{\bY}^{A})$.
\end{enumerate} 
\end{thm}

This theorem shows that {\em a posteriori} our knowledge about the underlying conditional distribution of $\bY$ given $\bX$ can again be represented by the same two-stage procedure that randomly partitions the predictor space and then generates the response distribution accordingly on each of the predictor blocks, except that now the parameters that characterize this two-stage procedure have been updated to reflect the information contained in the data. Moreover, the theorem also provides a recipe for computing these posterior parameters based on $\Phi(A)$ and $M(A)$. Given this exact posterior, Bayesian inference can then proceed---samples can be drawn from the posterior cond-OPT directly through vanilla Monte Carlo (as opposed to MCMC) and summary statistics calculated.

In the next section, we provide more details on how to implement such inference in practice. Before that, we present our last theoretical result about the cond-OPT prior---its posterior consistency, which assures the statistician that the posterior cond-OPT distribution will ``converge'' in some sense to the truth as the amount of data increases. To this end, we first need a notion of neighborhoods for conditional
densities under which such convergence holds. We adopt the notion discussed in \cite{pati:2011} and
\cite{norets:2014}, by which a (weak) neighborhood of a conditional density function is defined in terms of a (weak) neighborhood of the corresponding joint density. More specifically, for a 
conditional density function 
$f_0(\cdot|\cdot):\om_{\bY}\times \om_{\bX} \rightarrow [0,\infty)$, weak neighborhoods
with respect to a marginal density $f^0_{\bX}(\cdot)$ on $\om_{\bX}$ are collections of
conditional densities of the form
\[
U=\Bigl\{f(\cdot|\cdot) : \Big|\int g_i f(\cdot |\cdot)f^0_{\bX}d\mu - 
\int g_i f_0(\cdot|\cdot)f^0_{\bX}d \mu \Big| < \epsilon_i, i =
1,2,\ldots, l \Bigr\}
\]   
where the $g_i$'s are bounded continuous functions on $\om_{\bX} \times
\om_{\bY}$.
\begin{thm}[Weak consistency]
\label{thm:consistency}
 Let $(x_1, y_1), (x_2,y_2),\ldots$ be independent
identically distributed vectors from a probability distribution on
$\om_{\bX} \times \om_{\bY}$, $F$, with density $dF/d\mu=f(x,y)=f(y|x)f_{\bX}(x)$. Suppose the conditional density $f(\cdot|\cdot)$ is generated from a cond-OPT prior for which the conditions in Theorem~\ref{thm:large_support} all hold. In addition, assume that the conditional density function $f(\cdot|\cdot)$ and the joint density
$f(\cdot,\cdot)$ are bounded.
Then for any weak neighborhood of $f(\cdot|\cdot)$ w.r.t $f_{\bX}$, $U$, we have
\[
\pi(U|(x_1,y_1),(x_2,y_2),\ldots, (x_n,y_n)) \longrightarrow 1
\]
with $F^{\infty}$ probability 1, where
$\pi(\cdot|(x_1,y_1),(x_2,y_2),\ldots, (x_n,y_n))$ denotes the cond-OPT posterior for $f(\cdot|\cdot)$.
\end{thm}

\section{Practical implementation}
\label{sec:comp_issues}
Next we address some practical issues in computing the posterior and implementing the inference. For simplicity, from now on we shall refer to a set $A\subset \om_{\bX}$ that can arise during the (Stage~I) recursive partitioning procedure as a ``node'' (i.e., as a node in the partition tree).  

A prerequisite for applying Theorem~\ref{thm:conjugacy} is the availability of the $\Phi(A)$ terms, which can be determined recursively through Eq.~\eqref{eq:cond_int}. Of course, to carry out the computation of $\Phi(A)$ one must specify terminal conditions on Eq.~\eqref{eq:cond_int}, or in other words, on what kind of $A$'s the recursion should terminate. We call such nodes {\em terminal} nodes. 

There are two kinds of nodes for which the value of $\Phi(A)$ is available directly according to theory, and thus recursion can terminate on them. They are (i) nodes that cannot be further divided under the partition rule $\R_{\bX}$, and (ii) nodes that contain no more than one data
point.  For a node $A$ that cannot be further divided, we must have $\rho(A)=1$ and so $\Phi(A)=M(A)$. For a node $A$ with no data point, it has no contribution to the likelihood and so $\Phi(A)=1$. For a node $A$ with exactly one data point, $\Phi(A)$ is the predictive density of the local OPT on $A$ evaluated at that data point, which is exactly the density of the prior mean of the local OPT and is directly known when the default symmetric and self-similar prior specification for the local OPTs is adopted as recommended in \cite{wongandma:2010}. 
 
Note that with these two types of ``theoretical'' terminal nodes, in principle the recursion will eventually terminate if one divides the predictor space deep enough. In practice, however, it is unnecessary to take the recursion all the way down to these theoretical terminal nodes. Instead, one can adopt early termination by imposing a technical limit---such as a minimum size (or maximum depth) of the nodes either in terms of the natural measure $\mu_{\bX}(A)$ or the number of observations therein $n(A)$---to end the recursion. Nodes that are smaller than the chosen size threshold are forced to be terminal, which is equivalent to setting $\rho(A)=1$ and thus $\Phi(A)=M(A)$ for these nodes. We call these nodes ``technical'' terminal nodes.

With these theoretical and technical terminal nodes, one can then compute $\Phi(A)$ through the recursion formula $\eqref{eq:cond_int}$, and compute the posterior according to Theorem~\ref{thm:conjugacy}. Putting all the pieces together, we can summarize the procedure to carry out Bayesian inference with the cond-OPT prior as a four-step recipe: 
\begin{enumerate}
\item[I.] For all nodes (terminal or non-terminal), compute $M(A)$.
\vspace{0.2em}

\item[II.] For each non-terminal node $A$ (those that are ancestors of the terminal nodes), use Eq.~\eqref{eq:cond_int} to recursively compute $\Phi(A)$.
\vspace{0.2em}

\item[III.] Given the values of $M(A)$ and $\Phi(A)$, apply Theorem~\ref{thm:conjugacy} to get the parameter values of the posterior cond-OPT distribution.
\vspace{0.2em}

\item[IV.] Sample from the exact posterior by direct simulation of the random two-stage procedure, and/or compute summary statistics of the posterior.
\vspace{0.2em}

\end{enumerate}
For the last step, direct simulation from the posterior is straight-forward, but we have not discussed what summary statistics to compute and how to do that. This is problem-specific and will be illustrated in our numeric examples in \ref{sec:examples}.

\section{Examples}
\label{sec:examples}
In this section we provide four examples to illustrate inference using the
cond-OPT prior. The first two illustrate the estimation of conditional
densities, the latter two are for model selection and hypothesis testing. In these examples, the partition rules used on both $\om_{\bX}$ and $\om_{\bY}$ are always the coordinate-wise dyadic mid-split rule. We adopt the same prior specification across all the examples: the prior stopping probability on each non-terminal node is
always set to 0.5, the prior partition selection probability is always evenly spread over the possible ways to partition each set, and the probability assignment pseudo-counts for the local OPTs are all
set to 0.5. For continuous sample spaces, nodes at 12 levels down the partition tree, i.e., with $\mu_{\bX}(A)=\mu(\om_{\bX})/2^{12}$, are set to be the technical terminal nodes.

\begin{exam}[Estimating conditional density with abrupt changes over predictor values]
\label{ex:clear_boundary}
In this example we simulate $(X,Y)$ pairs according to the following
distributions.
\begin{align*}
X &\sim {\rm Beta}(2,2)\\
Y|X<0.25 &\sim {\rm Beta}(30,20)\\
Y|0.25 \leq X \leq 0.5 & \sim {\rm Beta}(10,30)\\
Y|X>0.5 &\sim {\rm Beta}(0.5,0.5).
\end{align*}
We generate data sets of three different sample sizes, $n=100$,
$n=500$, and $n=2,500$, and place the cond-OPT prior on the
distribution of $Y$ given $X$. Following the four-step recipe given in the previous section, we can compute the posterior cond-OPT and sample from it. 

A representative summary of the posterior partitioning mechanism is the so-called hierarchical {\it maximum a posteriori} (hMAP) \cite{wongandma:2010} partition tree, which can be computed from the posterior analytically \cite{wongandma:2010} and is plotted in \ref{fig:density_tree} for the different sample sizes. (Chipman et al \cite{chipman:1998} and Wong and Ma \cite{wongandma:2010} both discussed reasons why the commonly adopted MAP is not a good summary for tree-structured posteriors due to their multi-level nature. See \cite[Sec.~4.2]{wongandma:2010} for further details and reasons why the hMAP is often preferred to the MAP.) 

In \ref{fig:density_tree}, within each ``leaf'' node we plot the corresponding posterior mean of the
local OPT. Also plotted for each node is
the posterior stopping probability. Even with only 100 data points, the posterior
suggests that $\om_{\bX}$ should be divided into three
pieces---[0,0.25], [0.25,0.5], and [0.5,1]---within which
the conditional distribution of $Y|X$ is homogeneous across $X$. Note
that the posterior stopping probabilities on
those three intervals are large, in contrast to the near 0 values on the larger
sets. Reliably estimating the actual conditional density
function on these sets nonparametrically appears to require more than 100 data points. In this example, a sample size
of 500 already does a decent job. 

\begin{figure}[ht!]
  \begin{center}
    \mbox{
      \subfigure[$n=100$]{\includegraphics[width=11cm,height=5cm,clip=true,trim=10mm
        50mm 30mm 20mm]{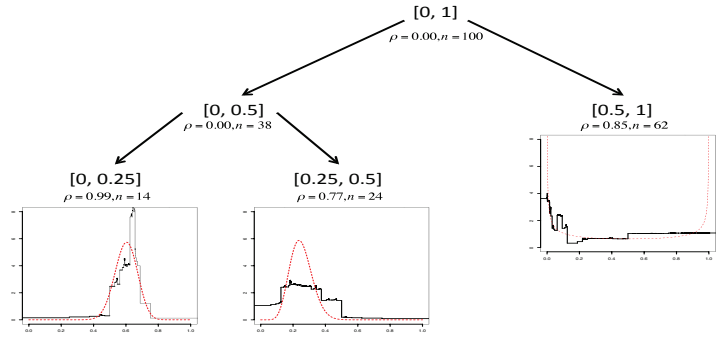}}
    }

    \mbox{
      \subfigure[$n=500$]{\includegraphics[width=11cm,height=5cm, clip=true,trim=10mm
        50mm 30mm 20mm]{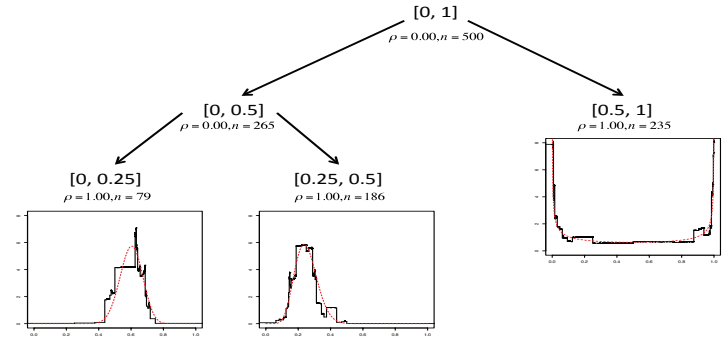}}
    }

    \mbox{
      \subfigure[$n=2500$]{\includegraphics[width=11cm,height=5cm, clip=true,trim=10mm
        50mm 30mm 20mm]{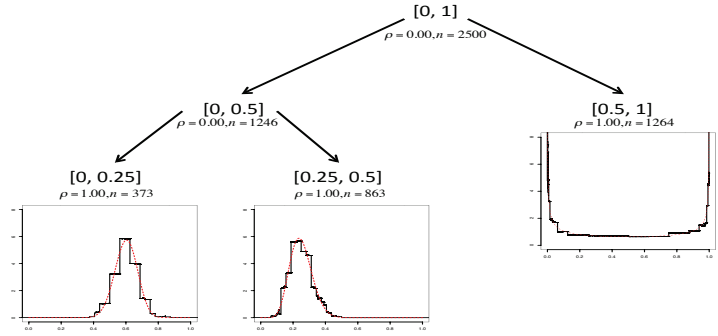}}
    }

    \caption{The hMAP partition tree structures on $X$ and the posterior
      mean estimate of $Y|X$ conditional on the random partition for
      Example~\ref{ex:clear_boundary}. For each node, $\rho$ indicates the posterior stopping probability for each node and $n$ represents the number of data points in each node. The plot under each stopped node gives the mean of the posterior local OPT for $Y$ within that node (solid line) along with the true conditional densities (dashed line).} 

    \label{fig:density_tree}
  \end{center}
\end{figure}

We compare both the model fit and the computing speed of our cond-OPT prior to two existing Bayesian nonparametric models for conditional densities---namely the linear dependent Dirichlet process mixture of normals (LDDP) \cite{deiorio:2009} and the linear dependent Dirichlet process mixture of Bernstein polynomials (LDBP) \cite{barrientos:2017}, both available in the {\tt DPpackage} in {\tt R}. In this example and the next, for LDDP and LDBP, we draw 1,000 posterior samples from the MCMC with a 2,000 burn-in period and a thinning interval of 3, and used prior specification given in the examples of the DPpackage. For details, please see the documentation for these two functions in the {\tt DPpackage} manual on {\tt CRAN}.

To evaluate model fit, we generate an additional testing data set from the true distribution of $(X,Y)$, and calculate the log-$p$ score (i.e., the log predictive likelihood of the testing set) for the three methods.  \ref{tab:logp_ex1} presents the log-$p$ score for the three methods from a typical simulated data set and the corresponding computing time on the same laptop computer with an Intel Core-i7 CPU using a single core without parallelization. A surprising phenomenon is that the performance of LDBP, in terms of the log-$p$ score for the testing sample, is not always monotone increasing in the sample size---that is, a larger training sample does not always lead to better fit on the testing set. In the particular simulation reported in \ref{tab:logp_ex1}, the preformance of LDBP is actually monotone decreasing with sample size. The cause for this is likely to be that under those models the conditional density is assumed to be smoothly varying over the predictors, and so as the true conditional density involves abrupt changes, the misspecified models can be consistently wrong even with large sample sizes.
\begin{table}[h]
\caption{Log predictive score and computing time for three Bayesian nonparametric models on a simulated data set in Example~\ref{ex:clear_boundary}}
\begin{center}
\begin{tabular}{c|ccc|ccc|ccc}
\hline
&&$n=100$ &&&$n=500$ &&& $n=2500$&\\
&cond-OPT & LDDP & LDBP&cond-OPT & LDDP & LDBP&cond-OPT & LDDP & LDBP\\
\hline
log-$p$ &75.5&17.6&34.3&78.2&24.9&31.4&81.5&33.1&27.8\\
CPU time (s) &0.48&$7.3\!\times\! 10^2$&$1.3\!\times\! 10^2$&0.82&$3.4\!\times\! 10^3$&$4.0\!\times\! 10^2$&1.8&$1.8\!\times\! 10^4$ &$1.7\!\times\! 10^3$\\
\hline
\end{tabular}
\end{center}
\label{tab:logp_ex1}
\end{table}%
\end{exam}

The previous example favors our method because (1) there are a 
small number of clear boundaries of change for the underlying 
conditional distribution,  and to a lesser extent (2) those boundaries---namely 0.25 and 0.5---lie on the
potential partition points of the partition rule. In the next example, we
examine the case in which the conditional
distribution changes smoothly across a continuous $X$ without any boundary
of abrupt change.

\begin{exam}[Estimating conditional densities that vary smoothly with predictor values]
\label{ex:bvnorm}
In this example we generate $(X,Y)$ from a bivariate normal
distribution.
\[ (X,Y)' \sim {\rm BN}\Biggl(\begin{pmatrix}
  0.6 \\ 0.4 \end{pmatrix},\begin{pmatrix} 0.1^2 & 0.005 \\ 0.005 &
  0.1^2\\\end{pmatrix}\Biggr).\]
We generate a data set of size $n=2,000$, and apply the cond-OPT prior
on the distribution of $Y$ given $X$ as we did in the previous
example. Again we compute the posterior cond-OPT following our four-step recipe. The hMAP tree and the posterior mean estimate
of the conditional density given the random partition is presented in \ref{fig:density_tree2}.   
Because the underlying predictor space $\om_{X}$ is unbounded, for simplicity in the above we used the empirically observed range of $X$ as $\om_X$, which happens to be $\om_X=[0.24, 0.92]$ for our simulated example. (Other ways to handle this situation include transforming $X$ to have a compact support such as through a CDF or rank transform.

One interesting observation is that the ``leaf'' nodes in \ref{fig:density_tree2} have
very large (close to 1) posterior stopping probability. This may seem
surprising as the underlying conditional distribution
is not the same for any neighboring values of $X$. The large
posterior stopping probabilities indicate that on those sets, where the sample size is not large, the gain in achieving better
estimate of the common features of the conditional distribution for
nearby $X$ values outweighs the loss in ignoring the difference
among them.  

\begin{figure}[h]
\vspace{-1em}
\begin{center}
    \leavevmode 
    \includegraphics[width=48em, clip=true,trim=30mm
        80mm 20mm 30mm]{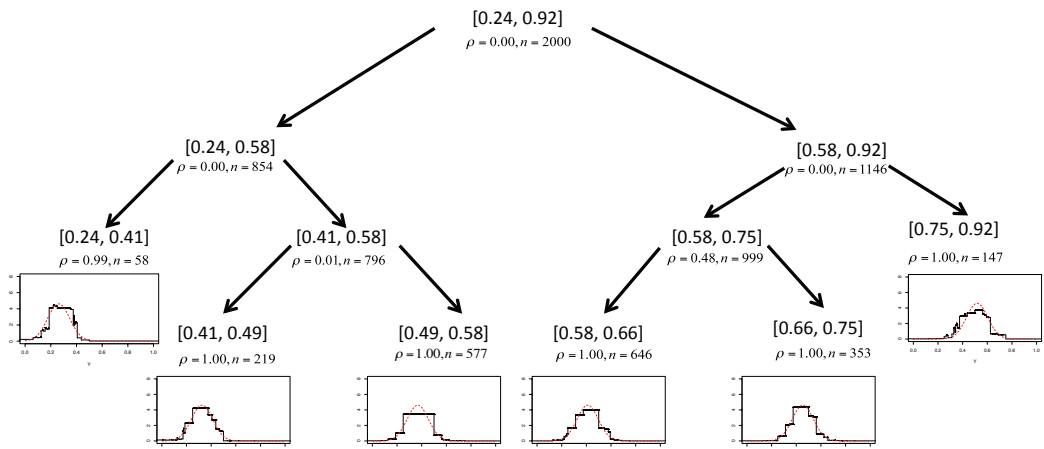}
     \caption{The hMAP tree on $\om_{X}$ and the predictive conditional density of $Y|X$ within the stopped sets conditional on the partition tree for a sample of size $n=2000$ in Example~\ref{ex:bvnorm}. The plot under each stopped node gives the mean of the posterior local OPT for $Y$ within that node (solid line) along with the true conditional densities at the center value of the stopped predictor intervals (dashed line). The $\rho$ label above each node is the posterior stopping probability for each node and $n$ represents the number of data points in each node.}
  \label{fig:density_tree2}
\end{center}
\end{figure}

Again, to compare the model fit and  computational efficiency with LDDP and LDBP, we repeat a set of simulations with different sample sizes $n=100$, 500, and 2500, and again use the log-$p$ score on a testing sample of size 100 to evaluate the performance. The results are summarized in \ref{tab:logp_ex2}, and they mostly confirm our intuition---the smooth priors overall outperform our model, especially for small sample sizes. The performance difference vanishes as the sample size increases. 

\begin{table}[h]
\caption{Log predictive score and computing time for three Bayesian nonparametric models on a simulated data set in Example~\ref{ex:bvnorm}}
\begin{center}
\begin{tabular}{c|ccc|ccc|ccc}
\hline
&&$n=100$ &&&$n=500$ &&& $n=2500$&\\
&cond-OPT & LDDP & LDBP&cond-OPT & LDDP & LDBP&cond-OPT & LDDP & LDBP\\
\hline
log-$p$ &75.4& 103 &  102 & 86.4 & 104&  104 &103 & 105 & 105 \\ 
CPU time (s) &0.8&$5.3\times 10^2$&$1.3\times 10^2$&1.4&$2.5\times 10^3$&$3.4\times 10^2$&2.5&$1.4\times 10^4$ &$1.9\times 10^3$\\
\hline
\end{tabular}
\end{center}
\label{tab:logp_ex2}
\end{table}%
\end{exam}

\begin{exam}[Model selection over binary predictors]
\label{ex:bernoulli}
Next we show how one can use cond-OPT to carry out model selection---that is, when multiple predictors are present, identifying the ones that affect the conditional distribution of $Y$. Consider the case in which $\bX = (X_1,X_2,\ldots,X_{30}) \in \{0,1\}^{30}$ forming a Markov Chain:
 \[ X_1\sim {\rm Bernoulli}(0.5) \quad \text{and} \quad P(X_i =X_{i-1}| X_{i-1})=0.7 \] 
for $i = 2,3,\ldots,30$. Suppose the conditional distribution of a continuous response $Y$ is
\begin{align*}
Y \sim \left\{ \begin{array}{ll} {\rm Beta}(1,6) & \text{if $(X_{5},X_{20},X_{30}) = (1,0,1)$}\\
    {\rm Beta}(12,16) & \text{if $(X_{5},X_{20})=(0,1)$}\\
    {\rm Beta}(3,4) & \text{otherwise.}    \end{array}
\right.
\end{align*}
In other words, three predictors $X_{5}$, $X_{20}$ and $X_{30}$ impact the response in an interactive manner. Our interest is in recovering this underlying interactive structure (i.e. the ``model''). To illustrate, we simulate 500 data points from this scenario and place a cond-OPT prior on $Y|\bX$, and consider predictor partitions up to four levels deep. This is achieved by setting $\rho(A)=1$ for $A$ that arises after four steps of partitioning, and it allows us to search for models involving up to four-way interactions. We again carry out the four-step recipe to get the posterior and calculate the hMAP. The hMAP tree structure along with the predictive conditional density for $Y|\bX$ within each stopped set given the random partition is presented in \ref{fig:bernoulli_tree}. The posterior concentrates on partitions involving $X_{5}$, $X_{20}$ and $X_{30}$ out of the 30 variables. While the predictive conditional density for $Y|\bX$ is very rough given the limited number of data points in the stopped sets, the posterior recovers the exact interactive structure of the predictors with little uncertainty. 

\begin{figure}[t]
\begin{center}
    \leavevmode 
    \includegraphics[width=35em, clip=TRUE, trim = 12mm 55mm 15mm 22mm]{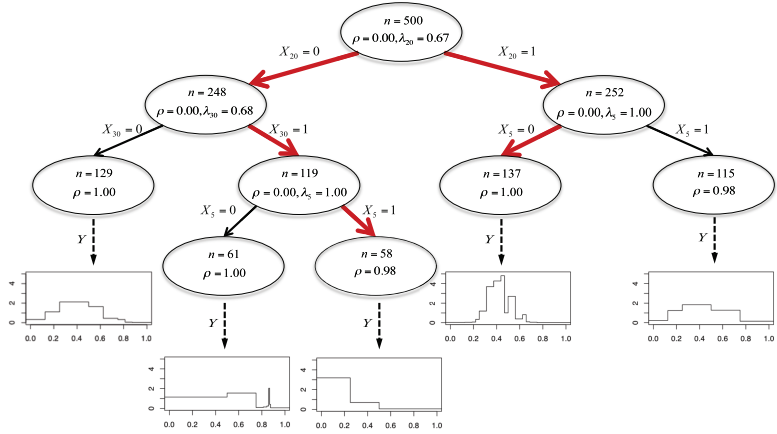}
     \caption{The hMAP tree structure on $\om_{\bX}$ and the posterior mean
       estimate of $Y|\bX$ given the random partition in each of the stopped sets for Example~\ref{ex:bernoulli}. The bold arrows indicate the ``true model''---predictor combinations that correspond to ``non-null'' $Y|\bX$ distributions. For each node, $\rho$ indicates the posterior stopping probability for each node, $\lambda$ represents the posterior selection probability for the most probable direction if the partition does not stop on the node, and $n$ represents the number of data points in each node.}
  \label{fig:bernoulli_tree}
\end{center}
\end{figure} 
\begin{figure}[h!]
  \begin{center}
    \mbox{
      \subfigure[$n=500$]{\includegraphics[width=8cm, clip=true,trim=0mm
        10mm 0mm 0mm]{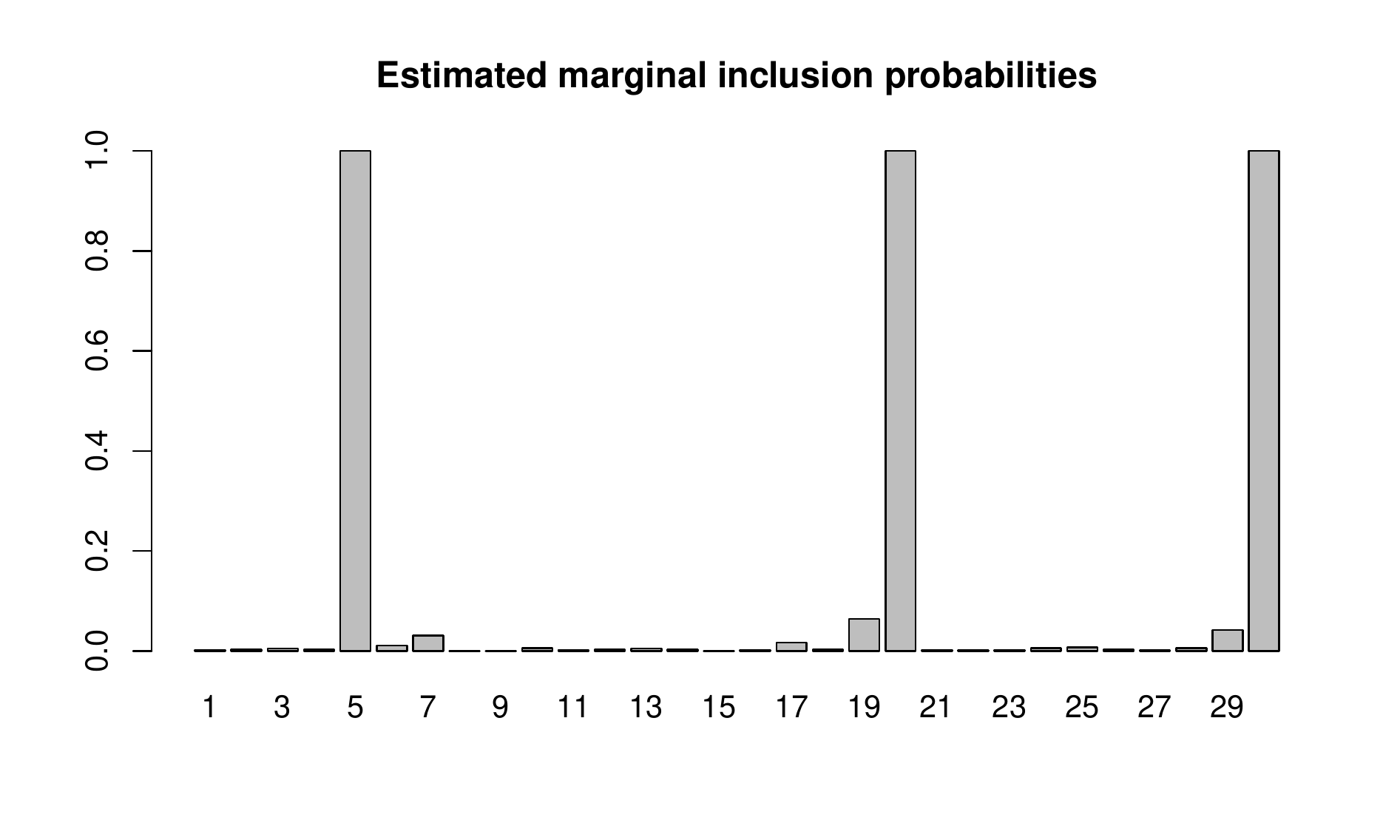}}
      \hspace{1em}
      \subfigure[$n=200$]{\includegraphics[width=8cm,clip=true,trim=0mm
        10mm 0mm 0mm]{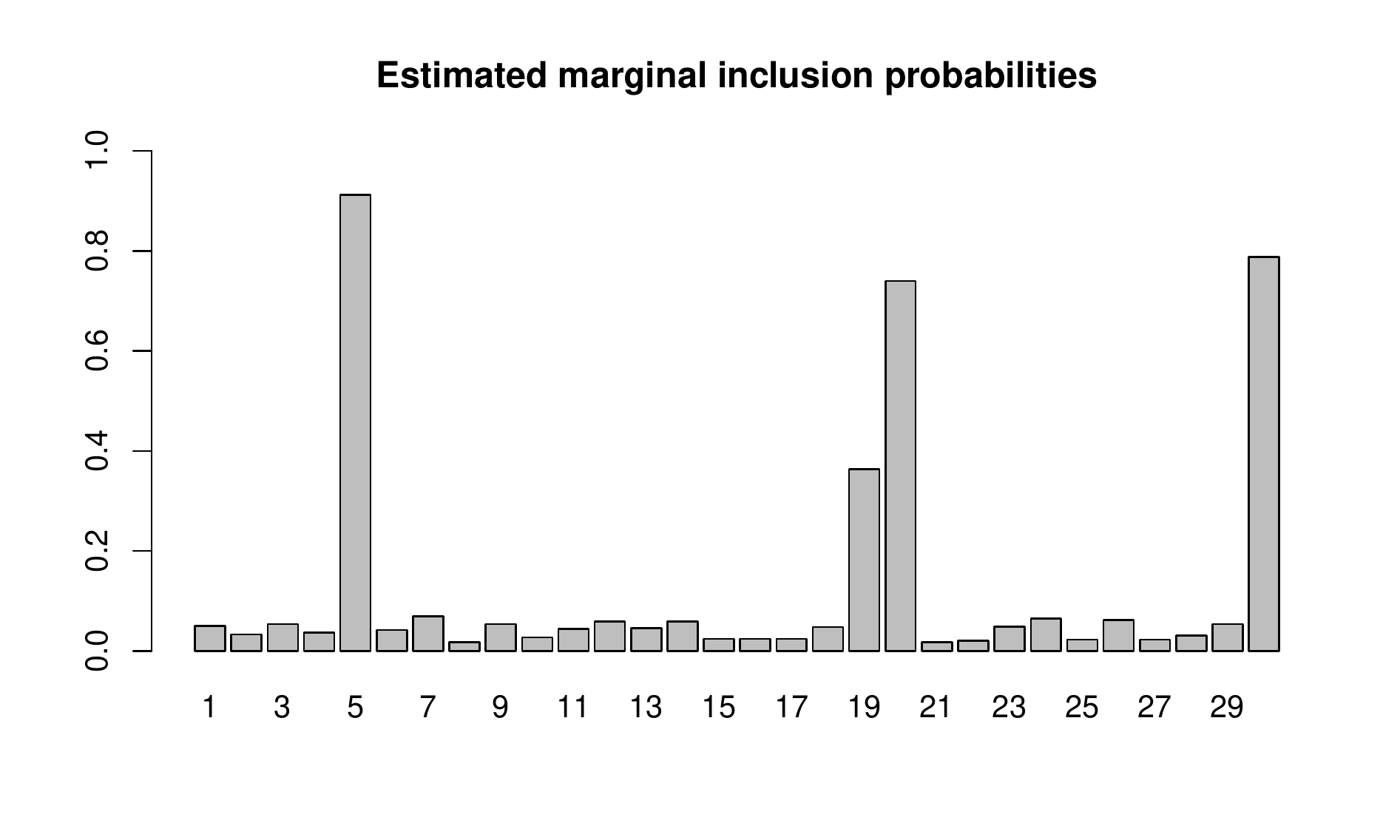}}
 
}
    \caption{Estimated posterior marginal inclusion probabilities for the 30 predictors in Example~\ref{ex:bernoulli} for two different sample sizes. The estimates are computed over 1,000 draws from the corresponding posteriors.} 
    \label{fig:marg_incl_probs}
  \end{center}
\end{figure}

In addition, we sample from the posterior and use the proportion of times each predictor appears in the sampled models to estimate the posterior marginal inclusion probabilities. Our estimates based on 1,000 draws from the posterior are presented in \ref{fig:marg_incl_probs}(a). Note that the sample size 500 is so large that the posterior marginal inclusion probabilities for the three relevant predictors are all close to 1 while those for the other predictors are close to 0. We carry out the same simulation with a reduced sample size of 200, and plot the estimated posterior marginal inclusion probabilities in \ref{fig:marg_incl_probs}(b). We see that with a sample size of 200, one can already use the posterior to reliably recover the relevant predictors.
\end{exam}

\begin{exam}[Test of independence]
In this example, we illustrate an application of the cond-OPT prior for hypothesis testing. In particular, we use it to test the independence between $\bX$ and $\bY$. To begin, note that $\rho(A|\bx,\by)$ in Theorem~\ref{thm:conjugacy} gives the posterior probability for the conditional distribution of $\bY$ to be constant over all values of $\bX$ in $A$, or in other words, for $\bY$ to be independent of $\bX$ on $A$. Hence, one can consider $\rho(\om_{\bX}|\bx,\by)$ as a score for the statistical significance of dependence between the observed variables. A permutation null distribution of this statistic can be constructed by randomly pairing the observed $\bx$ and $\by$ values, and based on this, permutation $p$-values can be computed for testing the null hypothesis of independence. 

To illustrate, we simulate $\bX=(X_1,X_2,\ldots,X_{10})$ for a sample of size 400 under the same Markov Chain model as in the previous example, and simulate a response variable $Y$ as follows.
\[
Y \sim \left\{ \begin{array}{ll} {\rm Beta}(4,4) & \text{if $(X_1,X_2,X_5) = (1,1,0)$}\\
    {\rm Beta}(0.5,0.5) & \text{if $(X_5,X_8,X_{10})=(1,0,0)$}\\
    {\rm Unif}[0,1] & \text{otherwise.}    \end{array}
\right.
\]
In particular, $\bY$ is dependent on $\bX$ but there is no mean or median shift in the conditional distribution of $Y$ over different values of $\bX$. \ref{fig:test_ind} gives the histogram of $\rho(\om_{\bX}|\bx,\by)$ for 1,000 permuted samples where the vertical dashed line indicates the $\rho(\om_{\bX}|\bx,\by)$ for the original simulated data, which equals 0.0384. For this particular simulation, 7 out of the 1,000 permuted samples produced a more extreme test statistic.
\begin{figure}[h!]
\begin{center}
    \leavevmode 
    \includegraphics[width=20em, clip=true, trim=0mm
        10mm 0mm 15mm ]{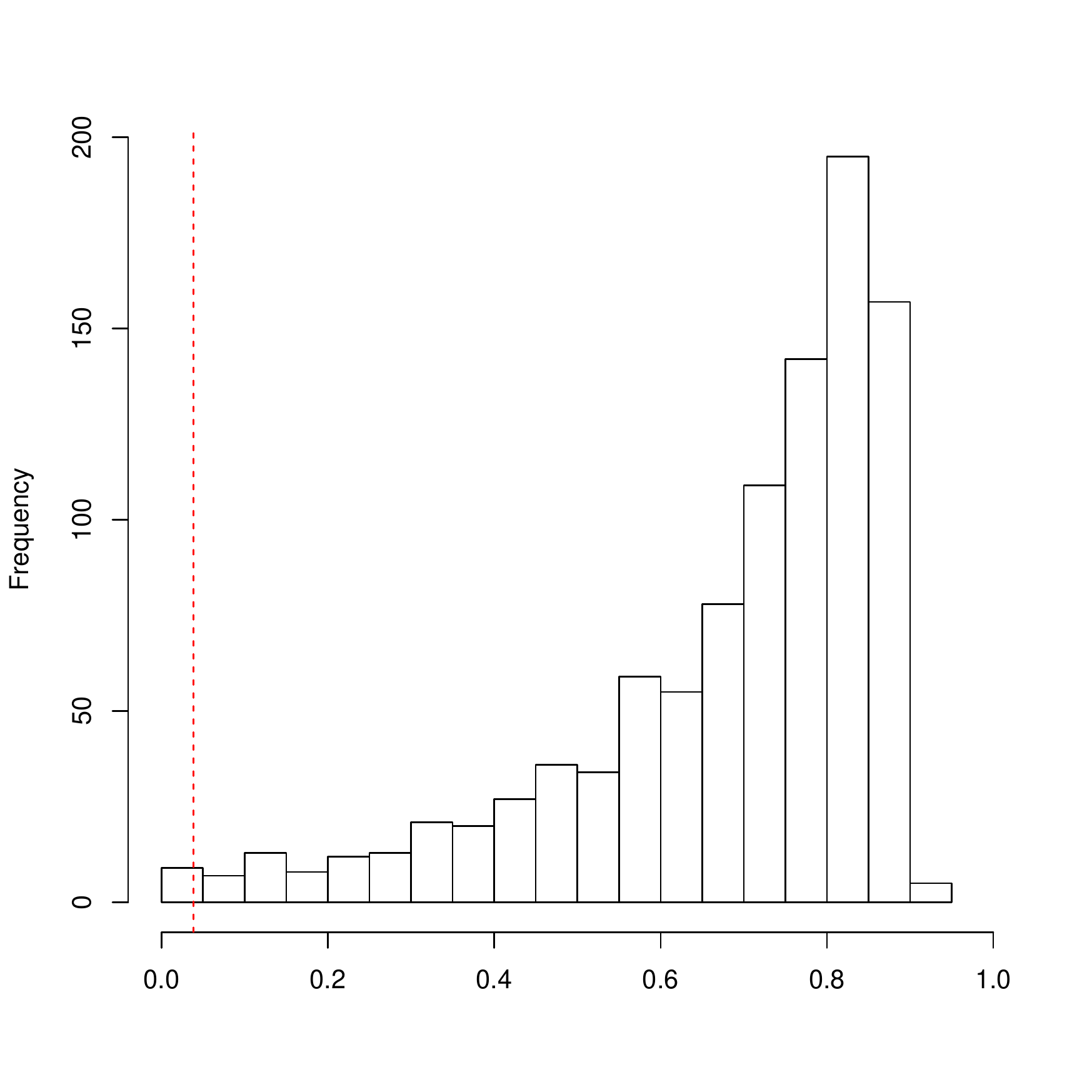}
     \caption{Histogram of $\rho(\om_{\bX}|\bx,\by)$ for 1,000 permuted samples. The vertical line indicates $\rho(\om_{\bX}|\bx,\by)$ for the original data.}
  \label{fig:test_ind}
\end{center}
\end{figure}

\noindent Remark I: Note that by symmetry one can place a cond-OPT prior on the conditional distribution of $\bX$ given $\bY$ as well and that will produce a corresponding posterior stopping probability $\rho(\om_{\bY}|\by,\bx)$. One can thus alternatively use $\min\{\rho(\om_{\bX}|\bx,\by),\rho(\om_{\bY}|\by,\bx)\}$ as the test statistic for independence.\\

\noindent Remark II: Testing using the posterior stopping probability $\rho(\om_{\bX}|\bx,\by)$ is equivalent to using a Bayes factor (BF). To see this, note that the BF for testing independence under the cond-OPT can be written as
\begin{align*}
{\rm BF}_{\bY|\bX} & = \frac{\sum_{j=1}^{N(A)}\lambda_j(A) \prod_{i}\Phi(A^j_i)}{M(A)}
\end{align*}
with $A=\om_{\bX}$ where the numerator is the marginal conditional likelihood of $\bY$ given $\bX$ if the conditional distribution of $\bY$ is not constant over $\bX$ (i.e.\ $\om_{\bX}$ is divided) and the denominator is that if the conditional distribution of $\bY$ is the same for all $\bX$ (i.e.\ $\om_{\bX}$ is undivided). By Eq.~\eqref{eq:cond_int} and Theorem~\ref{thm:conjugacy}, 
\[ {\rm BF}_{\bY|\bX} = \frac{\rho(\om_{\bX})}{1-\rho(\om_{\bX})} \left(\frac{1}{\rho(\om_{\bX}|\bx,\by)}-1\right),\]
which is in a one-to-one correspondence to $\rho(\om_{\bX}|\bx,\by)$ given the prior parameters.
\end{exam}

\section{Application to real data: multivariate conditional density estimation in flow cytometry}
\label{sec:flow}
In flow cytometry experiments for immunological studies, a number (typically 4 to 10) of biomarkers are measured on large numbers of blood cells. Estimated densities and conditional densities of such data can be used for tasks such as automatic classification of the cells \cite{malek:2014}. We apply cond-OPT to estimate the conditional density of markers ``CD4'' and ``CD8'' given two other markers ``FSC-H'' and ``FSC-W'' in a flow cytometry data set. So in this case both $\om_{\bX}$ and $\om_{\bY}$ are two-dimensional. This particular data set contains $n=455,472$ cells. Flow cytometry experiments often involve large numbers of cells, and thus practical methods must scale well in computing time and memory usage with respect to the number of observations. This poses great challenge to existing nonparametric models that require intense MCMC computation. The values of the four markers are measured in the range of [0,1]. We use maximum level of partitioning to 10 on both the predictor space $\om_{\bX}$ and the response space $\om_{\bY}$ but otherwise the same prior specification as before.

\begin{figure}[p]
\begin{center}
    \leavevmode 
    \includegraphics[width=54em]{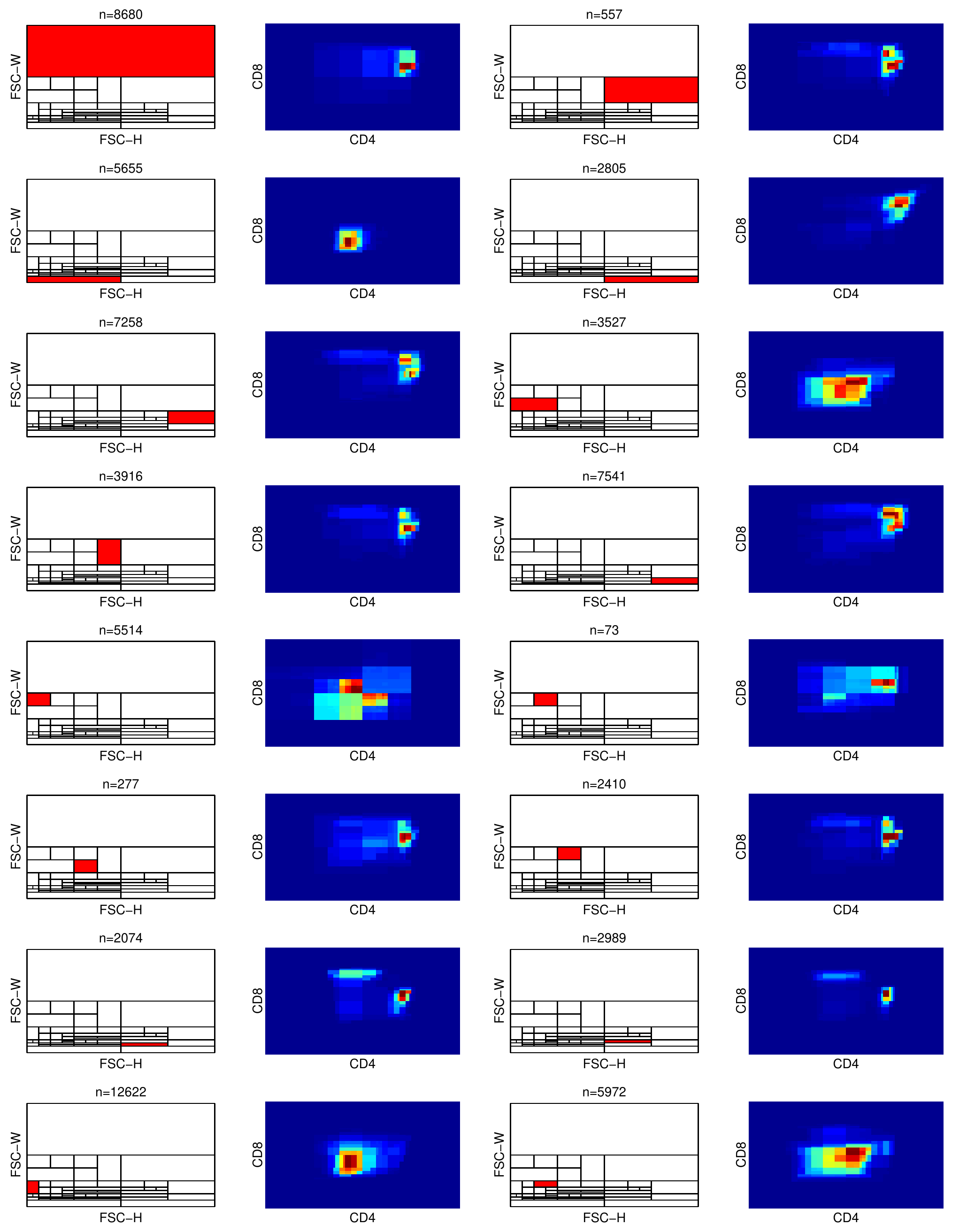}
     \caption{The posterior mean conditional densities of the two markers CD4 and CD8 given two other markers FSC-H and FSC-W conditional on the hMAP partition on FSC-H and FSC-W for the flow cytometry data set. The first and third columns indicate the corresponding predictor block (in red) in the hMAP partition with the number of observations labeled on top while the plots to their right illustrate the predictive conditional density on that block conditional on the random partition. Due to space constraint, we only show 16 out of the 50 predictor blocks.}
  \label{fig:flow}
\end{center}
\end{figure}

\ref{fig:flow} presents the posterior mean of the conditional density of CD4 and CD8 given FSC-H and FSC-W under the cond-OPT model given the random partition on the predictor space being the one induced under the hMAP tree, which splits the space into 50 pieces. A vast majority, in fact 44 out of the 50 predictor blocks are in fact not technical terminal regions, and so the model indeed smooths the conditional density over the predictor space. Because the number of predictor blocks is relatively large, we present the estimates for only 16 blocks in \ref{fig:flow}. The entire computation of the full posterior, the hMAP partition, as well as the conditional posterior expectation of the conditional density given the hMAP tree, took about 360 seconds to complete on a single 3.6GHz Intel Core-i7 3820 desktop core without parallelization and required about 8.2 Gbs of RAM. (Reducing the maximum level of partitions from 10 to 8 will reduce computing time to about 116 seconds and RAM to about 0.6 Gbs.)

\section{Discussion}
\label{sec:discussion}
In this work we have introduced a Bayesian nonparametric prior on the space
of conditional densities. This prior, which we call the conditional optional P\'olya tree, is
constructed based on a two-stage procedure that first divides the predictor
space $\om_{\bX}$ and then generates the conditional distribution of the response through local OPT processes. We have established several important theoretical properties of this
prior, namely large support, conjugacy and posterior consistency,
and have provided a practical recipe for Bayesian inference using this prior.

The construction of this prior does not depend on the marginal
distribution of $\bX$. One particular implication is that one can transform $\bX$
before applying the prior on $\bY|\bX$ without invalidating the posterior
inference. (Note that transforming $\bX$ is equivalent to choosing a
different partition rule on $\om_{\bX}$.) In certain situations it is
desirable to perform such a transformation on $\bX$. For example, if the
data points are very unevenly spread over $\om_{\bX}$, then some parts of the space may contain a very small number of data
points. There the posterior is mostly dominated by the prior
specification and does not provide much information about the
underlying conditional distribution. One way to mitigate this problem
is to transform $\bX$ so that the data are more evenly distributed over $\om_{\bX}$.
When $\om_{X}$ is one-dimensional, for example, this
can be achieved by a rank transformation on $X$. Another situation
in which a transformation of $\bX$ may be useful is when the
dimensionality of $\bX$ is very high. In this case a dimensionality
reduction transformation can be applied on $\bX$ before carrying out
the inference. Of course, in doing so one often loses the ability to
interpret the posterior conditional distribution of $\bY$ directly in
terms of the original predictors. An alternative approach when $\bX$ is high-dimensional is through variable selection that imposes certain sparsity assumptions, i.e., only a small number of predictors are affecting the conditional density. Exact calculation of full posterior and the marginal inclusion probabilities as we have carried out in Example~\ref{ex:bernoulli} is impractical when the number of predictors is large $(> 25 \sim 30)$. One strategy to overcome this difficulty is through sequential importance sampling as the one proposed in \cite{ma:2015}.

A general limitation of CART type randomized partitioning methods require a natural ordering of the space to be partitioned on. General partitioning strategies can be designed for unordered spaces, but then the computational efficiency of the proposed model would be lost.

Finally, we note that while we have used recursive partitioning in conjunction with the OPT to build a model for conditional density, one can build such models by replacing the OPT with other multi-scale density models in the family of P\'olya tree type models, such as the more recently introduced adaptive P\'olya tree (APT) \cite{ma:2016}. 

\section*{Software}
The proposed model has been implemented in the {\tt R} package {\tt PTT} (for P\'olya tree type models) as the function {\tt cond.opt}. A variant of the model that replaces the OPT with an APT is also implemented in the package as function {\tt cond.apt}. This package is currently available for download at \url{https://github.com/MaStatLab/PTT} and will be submitted to {\tt CRAN}.

\section*{Acknowledgment}
\label{sec:acknowledgment}
The flow cytometry data set was provided by EQAPOL (HHSN272201000045C), an
NIH/NIAID/DAIDS-sponsored, international resource that supports the development, implementation,
and oversight of quality assurance programs (Sanchez PMC4138253).

\section*{Appendix: Proofs}
\begin{proof}[Proof of Lemma~1]
The proof of this lemma is very similar to that of Theorem~1 in \cite{wongandma:2010}. 
 Let $T_1^{k}$ be the part of $\om_{\bX}$ that has not been stopped after $k$
levels of recursive partitioning. The random partition of $\om_{\bX}$ after $k$ levels of recursive partitioning can
be thought of as being generated in two 
steps. First suppose there is no stopping on any set and let $J^{*(k)}$ be
the collection of partition selection variables $J$ generated in the first $k$ levels of 
recursive partitioning. Let $\A^k(J^{*(k)})$ be the collection of sets
$A$ that arise in the first $k$ levels of non-stopping recursive
partitioning, which is determined by $J^{*(k)}$. Then we generate the stopping
variables $S(A)$ for each $A \in \A^k(J^{*(k)})$ successively for
$k=1,2,\ldots$, and once a set is stopped, let all its descendants be
stopped as well. Now for each $A \in \A^k(J^{*(k)})$, let $I^k(A)$ be the
indicator for $A$'s stopping status after $k$ levels of recursive
partitioning, with $I^k(A)=1$ if $A$ is not stopped and $=0$
otherwise. 
\begin{align*}
E(\mu_{\bX}(T_1^{k})|J^{*(k)}) &=
E\left(\sum_{A\in\A^k(J^{*(k)})}\mu_{\bX}(A)I^{k}(A)|J^{*(k)} \right)\\
&= \sum_{A\in\A^k(J^{*(k)})} \mu_{\bX}(A) E(I^{k}(A)|J^{*(k)})\\
&\leq \mu_{\bX}(\om_{\bX})(1-\delta)^k.
\end{align*}
Hence $E(\mu_{\bX}(T_1^{k})) \leq \mu_{\bX}(\om_{\bX})(1-\delta)^k$, by
Markov inequality and Borel-Contelli lemma, we have
$\mu_{\bX}(T_1^{k})\downarrow 0$ with probability 1. 
\end{proof}

\begin{proof}[Proof of Theorem~2]
We prove only the second result as the first follows by choosing
$f_{\bX}(x)\equiv 1/\mu_{\bX}(\om_{\bX})$. Also, we consider only the case when $\om_{\bX}$ and
$\om_{\bY}$ are both compact Euclidean rectangles, because the cases when
at least one of the two spaces is finite follow as simpler special
cases. For $x\in \om_{\bX}$ and 
$y \in \om_{\bY}$, let $f(x,y):=f_{\bX}(x)f(y|x)$ 
denote the joint density. First we assume that the joint density $f(x,y)$ is uniformly
continuous. In this case it is bounded on $\om_{\bX} \times \om_{\bY}$. We let $M
:= \sup f(x,y)$ and  
\[
\delta(\epsilon) := \sup_{|x_1-x_2|+|y_1-y_2|<\epsilon} |f(x_1,y_1)-f(x_2,y_2)|.
\]
By uniform continuity, we have
$\delta(\epsilon)\downarrow 0$ as $\epsilon \downarrow 0$. In
addition, we define
\begin{align*}
\delta_{\bX}(\epsilon)&:=\sup_{|x_1-x_2|<\epsilon}|f_{\bX}(x_1) - f_{\bX}(x_2)|\\
&\leq \int \sup_{|x_1-x_2|<\epsilon} |f(x_1,y)-f(x_2,y)|\mu_{\bY}(dy) \leq 
\delta(\epsilon)\mu_{\bY}(\om_{\bY}).
\end{align*}
Note that in particular the continuity of $f(x,y)$ implies the continuity of $f_{\bX}(x)$. 
Let $\sigma>0$ be any positive constant. Choose a positive constant $\epsilon(\sigma)$
such that $\delta_{\bX}(\epsilon(\sigma))=\delta(\epsilon(\sigma))\mu_{\bY}(\om_{\bY})<\max(\sigma/2,
\sigma^3/2)$. Because all the parameters in the cond-OPT are uniformly
bounded away 
from 0 and 1, there is positive probability that $\om_{\bX}$ will be
partitioned into $\om_{\bX} = \cup_{i=1}^{K} B_i$ where the diameter of
each $B_i$ is less than $\epsilon(\sigma)$, and the partition stops on each of
the $B_i$'s. (The existence of such a partition follows from the fine
partition criterion.) Let $A_i = B_i \cap \{\bX: f_{\bX}(x)\geq 
\sigma\}$, $P(\bX \in A_i) = \int_{A_i} f_{\bX}(x)\mu_{\bX}(dx)$, and
$f_i(y):=\int_{A_i}f(x,y)\mu_{\bX}(dx)/\mu_{\bX}(A_i)$ if $\mu_{\bX}(A_i)>0$, and
0 otherwise. Let $\mathcal{I} \subset \{1,2,\ldots,K\}$ be the set of
indices $i$ such that $\mu_{\bX}(A_i)>0$. Then  
\begin{align*}
&\quad \int |q(y|x) - f(y|x)|f_{\bX}(x)\mu(dx \times dy)\\
&\leq
\int_{f_{\bX}(x)<\sigma} |q(y|x) - f(y|x)|f_{\bX}(x)\mu(dx \times dy)\\ 
&+ \sum_{i\in \mathcal{I}}\int_{A_i \times \om_{\bY}}
\Big|q(y|x)-f_i(y)\cdot
\frac{\mu_{\bX}(A_i)}{P(\bX\in A_i)}\Big|f_{\bX}(x)\mu(dx \times dy)\\
& + \sum_{i\in \mathcal{I}}\int_{A_i \times \om_{\bY}}
f_i(y)\Big|\frac{\mu_{\bX}(A_i)}{P(\bX\in
  A_i)}-\frac{1}{f_{\bX}(x)}\Big|f_{\bX}(x) \mu(dx \times dy)\\
& +\sum_{i\in\mathcal{I}}\int_{A_i \times \om_{\bY}}\Big|f_i(y)-f(x,y)\Big|\mu(dx
\times dy).
\end{align*}
Let us consider each of the four terms on the right hand side in turn. First,
\[
\int_{f_{\bX}(x)<\sigma} |q(y|x) - f(y|x)|f_{\bX}(x)\mu(dx \times dy) \leq
2\sigma \mu_{\bX}(\om_{\bX}). 
\]
Note that for each $i\in \mathcal{I}$, $f_i(y)\mu_{\bX}(A_i)/P(\bX\in A_i)$
is a density function in $y$. Therefore by the large support
property of the OPT prior (Theorem~2 in \cite{wongandma:2010}), with
positive probability,
\[
\int_{\om_{\bY}}
\Big|q_{\bY}^{0,B_i}(y)-f_i(y)\cdot
\frac{\mu_{\bX}(A_i)}{P(\bX\in A_i)}\Big|\mu_{\bY}(dy) < \sigma,
\]
and so
\[
\int_{A_i \times \om_{\bY}}
\Big|q(y|x)-f_i(y)\cdot
\frac{\mu_{\bX}(A_i)}{P(\bX\in A_i)}\Big|f_{\bX}(x)\mu(dx \times dy) < \sigma P(\bX\in A_i)
\]
for all $i\in \mathcal{I}$. Also, for any $x \in A_i$, by the choice of $\epsilon(\sigma)$,
\[
\Big|\frac{\mu_{\bX}(A_i)}{P(\bX \in A_i)} - \frac{1}{f_{\bX}(x)}\Big| \leq
\frac{\delta_{\bX}(\epsilon(\sigma))}{\sigma(\sigma-\delta_{\bX}(\epsilon(\sigma))}\leq
\frac{\sigma^3/2}{\sigma^2/2}=\sigma. 
\]
Thus
\[
\int_{A_i \times \om_{\bY}}
f_i(y)\Big|\frac{\mu_{\bX}(A_i)}{P(\bX\in
  A_i)}-\frac{1}{f_{\bX}(x)}\Big|f_{\bX}(x) \mu(dx \times dy)\leq \sigma M \mu_{\bY}(\om_{\bY})P(\bX \in A_i).
\]
Finally, again by the choice of $\epsilon(\sigma)$,
$|f_i(y)-f(x,y)|\leq \delta(\epsilon(\sigma)) < \sigma$, and so
\[
\int_{A_i \times \om_{\bY}}\Big|f_i(y)-f(x,y)\Big|\mu(dx
\times dy) < \sigma \mu_{\bY}(\om_{\bY})\mu_{\bX}(A_i).
\]
Therefore for any $\tau>0$, by choosing a small enough $\sigma$, we can have 
\[ 
\int |q(y|x) - f(y|x)|f_{\bX}(x)\mu(dx \times dy) < \tau
\]
with positive probability. This completes the proof of the theorem for
continuous $f(x,y)$. Now we can approximate any density function
$f(x,y)$ arbitrarily close in $L_1$ distance by a continuous one
$\tilde{f}(x,y)$. The theorem still holds
because
\begin{align*}
\int |q(y|x) - f(y|x)|f_{\bX}(x)\mu(dx \times dy) & \leq \int q(y|x) |f_{\bX}(x)
- \tilde{f}_{\bX}(x)| \mu(dx \times dy)\\ 
&+ \int |q(y|x) - \tilde{f}(y|x)|\tilde{f}_{\bX}(x)\mu(dx \times dy)\\
&+ \int |\tilde{f}(x,y)-f(x,y)|\mu(dx \times dy).\\
&\leq \int |q(y|x) - \tilde{f}(y|x)|\tilde{f}_{\bX}(x)\mu(dx \times dy)\\
&+ 2\int |\tilde{f}(x,y)-f(x,y)|\mu(dx \times dy),
\end{align*}
where $\tilde{f}_{\bX}(x)$ and $\tilde{f}(y|x)$ denote the corresponding
marginal and conditional density functions for $\tilde{f}(x,y)$.
\end{proof}

\begin{proof}[Proof of Theorem~3]
Given that a set $A$ is reached during the random partitioning steps on
$\om_{\bX}$, $\Phi(A)$ is the marginal conditional likelihood of 
\[
\text{\{$\bY(A)=\by(A)$\} given \{$\bX(A)=\bx(A)$\}}.
\]
The first term on the right hand side of Eq.~(3.2), $\rho(A)M(A)$, 
is the marginal conditional likelihood of
\[
\text{\{Stop partitioning on $A$, $\bY(A)=\by(A)$\} given \{$\bX(A)=\bx(A)$\}.}
\]
Each summand in the second term, $(1-\rho(A))\lambda_j(A)\prod_i \Phi(A^j_i)$, is the
marginal conditional likelihood of
\[
\text{\{Partition $A$ in the $j$th way, $\bY(A)=\by(A)$\} given \{$\bX(A)=\bx(A)$\}.}
\]
Thus the conjugacy of the prior and the posterior updates for $\rho$,
$\lambda_j$ and OPT$(\R^A_{\bY};\rho_{\bY}^A,\blam_{\bY}^A,\balp_{\bY}^A)$ follows from
Bayes' Theorem and the posterior conjugacy of the standard optional
P\'olya tree prior (Theorem~3 in \cite{wongandma:2010}).
\end{proof}

\begin{proof}[Proof of Theorem~4]
By Theorem~2.1 in \cite{norets:2014}, which follows directly from Schwartz's
theorem (see \cite{schw65} and \cite[Theorem~4.4.2]{ghosh03}), we just need to prove that the prior places 
positive probability mass in arbitrarily small Kullback-Leibler
(K-L) neighborhoods of $f(\cdot|\cdot)$ w.r.t $f_{\bX}$. Here a K-L
neighborhood w.r.t $f_{\bX}$ is defined to be the collection of
conditional densities
\[
K_{\epsilon}(f) = \Bigl\{h(\cdot|\cdot) : \int
f(y|x)\log\frac{f(y|x)}{h(y|x)}f_{\bX}(x)\mu(dx \times dy) < \epsilon \Bigr\}
\]
for some $\epsilon>0$.

To prove this, we just need to show that any conditional density that
satisfies the conditions given in the theorem can be approximated
arbitrarily well in K-L distance by a piecewise constant conditional density of the sort that
arises from the cond-OPT procedure. We first assume
that $f(\cdot|\cdot)$ is continuous.
Following
the proof of Theorem~2, let $\delta(\epsilon)$
denote the modulus of continuity of $f(\cdot|\cdot)$.
Let $\om_{\bX}=\cup_{i=1}^K A_{i}$ be a reachable partition of $\om_{\bX}$ such
that the diameter of each partition block $A_i$ is less than $\epsilon$.
Next, for each $A_i$, let $\om_{\bY}=\cup_{j=1}^{N} B_{ij}$ be a partition
on $\om_{\bY}$ allowed under OPT$(\R_{\bY};\rho_{\bY}^{A_i},\blam_{\bY}^{A_i},\balp_{\bY}^{A_i})$ such that the diameter
of each $B_{ij}$ is also less than $\epsilon$. Let
\[
g_{ij} = \sup_{ x \in A_{i}, y \in B_{ij}} f(y|x)\quad \text{ and } \quad
g_i(y) = \sum_{j} g_{ij} I_{B_{ij}}(y).
\]
Let $G_i=\int_{A_i \times \om_{\bY}}  g_i(y) f_{\bX}(x) \mu(dx \times dy)$. Then
\begin{align*}
0 \leq \sum_i G_i - 1 &= \sum_i \int_{A_i \times \om_{\bY}} \bigl(g_i(y)-f(y|x)\bigr)f_{\bX}(x)d\mu 
\leq
\delta(2\epsilon)\mu_{\bY}(\om_{\bY}),
\end{align*}
and so 
$\sum_i G_i \leq 1 + \delta(2\epsilon)\mu_{\bY}(\om_{\bY})$. 

Now let $g(y|x) = \sum_{i} \left(g_i(y)/\int_{\om_{\bY}} g_i(\tilde{y})\mu_{\bY}(d\tilde{y})\right)I_{A_i}(x)$, which is a step
function that can arise from the cond-OPT prior. Then
\begin{align*}
0 & \leq \int f(y|x)\log\bigl(f(y|x)/g(y|x)\bigr)f_{\bX}(x)d\mu\\
&= \sum_{i}\Biggl(\int_{A_i \times \om_{\bY}}f(y|x)\log\bigl(f(y|x)/g_i(y)\bigr)f_{\bX}(x)d\mu \\& \qquad \quad +
\int_{A_i \times \om_{\bY}} f(y|x)\log\left( \int_{\om_{\bY}} g_i(\tilde{y})\mu_{\bY}(d\tilde{y})  \right)f_{\bX}(x)d\mu \Biggr)\\
& \leq \sum_{i} \log\left( \int_{\om_{\bY}} g_i(\tilde{y})\mu_{\bY}(d\tilde{y})  \right) P(\bX\in A_i) \\
& \leq \log\left(\sum_{i}\int_{\om_{\bY}} g_i(\tilde{y})\mu_{\bY}(d\tilde{y}) P(\bX\in A_i) \right) = \log(\sum_i G_i)
\leq \delta(2\epsilon)\mu_{\bY}(\om_{\bY}),
\end{align*}
which can be made arbitrarily close to 0 by choosing a small enough
$\epsilon$. Now if $f(\cdot |\cdot)$ is not continuous, then for any $\epsilon'>0$, there
exists a compact set $E \subset \om_{\bX} \times \om_{\bY}$ such that $f(\cdot|\cdot)$
is uniformly continuous on $E$ and $\mu(E^c)<\epsilon'.$ Now let
\[
g_{ij} = \left(\sup_{ (x,y) \in E\cap ( A_{i} \times B_{ij}) } f(y|x) +\delta(\epsilon/2)\right) \vee \epsilon''
\]
for some constant $\epsilon''>0$, while keeping the definitions of
$g_{i}$, $G_{i}$ and $g(y|x)$ in terms of
$g_{ij}$ unchanged. Let $M$ be a finite upperbound of
$f(\cdot|\cdot)$ and $f(\cdot,\cdot)$. We have  
\begin{align*}
\sum_i G_i - 1 &= \sum_i \int_{E\cap ( A_i  \times \om_{\bY})}
\bigl(g_i(y)-f(y|x)\bigr)f_{\bX}(x)d\mu \\
& \qquad + \sum_i\int_{E^c \cap ( A_i  \times \om_{\bY}) } \bigl(g_i(y)-f(y|x)\bigr)f_{\bX}(x)d\mu.
\end{align*}
Thus,
\[
\sum_i G_i - 1 \geq \delta(\epsilon/2) \mu_{\bY}(\om_{\bY}) - (2M+\epsilon'')M\mu_{\bY}(\om_{\bY})\epsilon',
\]
which is positive for small enough $\epsilon'$.
At the same time,
\begin{align*}
\sum_i G_i - 1 \leq \bigl(\delta(2\epsilon)+\epsilon''\bigr)\mu_{\bY}(\om_{\bY}) + (2M+\epsilon'')M\mu_{\bY}(\om_{\bY})\epsilon',
\end{align*}
which can be made arbitrarily small by taking
$\epsilon$, $\epsilon'$, and $\epsilon''$ all $\downarrow 0$. 

Now 
\begin{align*}
0 & \leq \int f(y|x)\log\bigl(f(y|x)/g(y|x)\bigr)f_{\bX}(x)d\mu\\
&= \sum_{i}\Biggl(\int_{A_i \times \om_{\bY}}f(y|x)\log\bigl(f(y|x)/g_i(y)\bigr)f_{\bX}(x)d\mu \\ & \hspace{1em}  +
\int_{A_i \times \om_{\bY}} f(y|x)\log\left( \int_{\om_{\bY}} g_i(\tilde{y})\mu_{\bY}(d\tilde{y})  \right)f_{\bX}(x)d\mu \Biggr)\\
&=\sum_{i}\int_{E \cap (A_i \times
  \om_{\bY})}f(y|x)\log\bigl(f(y|x)/g_i(y)\bigr)f_{\bX}(x)d\mu\\
&\hspace{1em} + \sum_{i}\int_{E^c
  \cap (A_i \times \om_{\bY})}
f(y|x)\log\bigl(f(y|x)/g_i(y)\bigr)f_{\bX}(x)d\mu \\
 & \hspace{1em} + \sum_{i}\int_{A_i \times \om_{\bY}} f(y|x)\log\left( \int_{\om_{\bY}} g_i(\tilde{y})\mu_{\bY}(d\tilde{y})  \right)f_{\bX}(x)d\mu\\ 
&\leq 0 + M\epsilon'\log(M/\epsilon'')+\log(\sum_i G_i)\\
&\leq M\epsilon'\log(M/\epsilon'')+\bigl(\delta(2\epsilon)+\epsilon''\bigr)\mu_{\bY}(\om_{\bY}) + (2M+\epsilon'')M\mu_{\bY}(\om_{\bY})\epsilon'.
\end{align*}
The right hand side $\downarrow 0$ if we take $\epsilon \downarrow 0$
and $\epsilon'=\epsilon''\downarrow 0$. This completes the proof.
\end{proof}

\bibliography{condOPT}

\end{document}